\documentstyle[12pt]{article}
\begin{document}
\begin{titlepage}
\begin{flushright}
NTUA--05/01 \\ hep-ph/0106033
\end{flushright}
\vspace{1cm}

\begin{centering}
\vspace{.4in} {\Large {\bf Soft Supersymmetry Breaking due to
Dimensional\\ \vspace{.3cm} Reduction over Non-Symmetric Coset Spaces.}}\\
\vspace{1.5cm}

{\bf P.~Manousselis}$^{a}$ and {\bf G.~Zoupanos}$^{b}$\\
\vspace{.2in} Physics Department, National Technical University,
\\ Zografou
Campus, 157 80 Athens, Greece.\\

\vspace{1.0in}

{\bf Abstract}\\

\vspace{.1in} A ten-dimensional supersymmetric $E_8$ gauge theory
is compactified over six-dimensional coset spaces, establishing
further our earlier conjecture that the resulting four dimensional
theory is a softly broken supersymmetric gauge theory in the case
that the used coset space is non-symmetric. The specific
non-symmetric six-dimensional spaces examined in the present study
are $Sp(4)/(SU(2) \times U(1))_{non-max.}$ and $ SU(3)/U(1) \times
U(1)$.
\end{centering}
\vspace{4.7cm}

\begin{flushleft}
$^{a}$e-mail address: pman@central.ntua.gr. Supported by
$\Gamma\Gamma$ET  grand 97E$\Lambda$/71.
\\ $^{b}$e-mail address:
George.Zoupanos@cern.ch. Partially supported by EU under the RTN
contract HPRN-CT-2000-00148 and the A.v.Humboldt Foundation.
\end{flushleft}
\end{titlepage}
\section{Introduction}
 Supersymmetry has been one of the essential ingredients of most
unification frameworks examined during the few last decades. This
is not a surprising fact given that the hope of understanding in a
unified manner particles with different spins and the aim that
such a unified description should be free of ultraviolet
divergencies have been in the core of most attempts. Supersymmetry
by definition points to a fulfillment of the first hope, while
already the first of the non-renormalization theorems in
supersymmetric theories \cite{Wess} guarantees improved
ultraviolet properties of such theories. On the other hand the
lack of any obvious sign of supersymmetry in the low energy
physics that have been explored during the last decades, has risen
the question of supersymmetry breaking to a fundamental issue
comparable to the existence of supersymmetry itself.

 Since the early days of supersymmetry several mechanisms such as the
Fayet-Iliopoulos \cite{Fayet}, the Fayet-O'Raifeartaigh
\cite{O'Raifeartaigh} have been proposed, while the celebrated
MSSM has been supplemented with a soft supersymmetry breaking
(SSB) sector which was supposed to be inherited to the low
energies by supergravity \cite{Cremmer}.

Concerning higher dimensional supersymmetric theories, like those
resulting in the field theory limit of superstrings, mostly two
mechanisms have been employed. One assumes that N=1 is preserved
by the compactification process and supersymmetry breaking has its
origin in the gaugino condensation taking place in the "hidden"
sector of the theory which eventually is communicated to the
observed sector. The other mechanism, called Scherk-Schwarz
\cite{Scherk-Schwarz} breaks supersymmetry in the process of
compactification. In ref.\cite{Pman} a new  mechanism, based on
the Coset Space Dimensional Reduction (CSDR)
\cite{Manton},\cite{Review},\cite{Kuby} has been proposed as the
possible origin of the SSB sector of a four dimensional
supersymmetric theory.

Specifically in ref.\cite{Pman} a ten-dimensional supersymmetric
gauge theory based on the group $E_{8}$ was reduced over the
six-dimensional non-symmetric coset space $G_{2}/SU(3)$ leading to
an $E_{6}$ softly broken supersymmetric GUT in four dimensions. On
the contrary the original supersymmetry of the theory was
completely broken by the dimensional reduction procedure over the
six-sphere $SO(7)/SO(6)$ which is a symmetric coset space. The
conjecture of ref.\cite{Pman} was that the above findings have a
wider validity. In the present work we establish further the
conjecture of ref.\cite{Pman} that dimensional reduction over
non-symmetric coset spaces leads {\it automatically} to softly
broken supersymmetric four-dimensional theories, by studying the
dimensional reduction of a ten-dimensional supersymmetric $E_{8}$
gauge theory over the rest two existing non-symmetric
six-dimensional coset spaces. We find that the dimensional
reduction over the non-symmetric coset spaces $Sp(4)/(SU(2) \times
U(1))_{non-max.}$ and $SU(3)/U(1) \times U(1)$ leads to softly
broken supersymmetric gauge theories in four dimensions with a
complete SSB sector, while no other term that could possibly spoil
the ultraviolet properties of the theories appears.
\section{The Coset Space Dimensional Reduction.}
Given a gauge theory defined in higher dimensions the obvious way
to dimensionally reduce it is to demand that the field dependence
on the extra coordinates is such that the Lagrangian is
independent of them. A crude way to fulfill this requirement is to
discard the field dependence on the extra coordinates, while an
elegant one is to allow for a non-trivial dependence on them, but
impose the condition that a symmetry transformation by an element
of the isometry group $S$ of the space formed by the extra
dimensions $B$ corresponds to a gauge transformation. Then the
Lagrangian will be independent of the extra coordinates just
because it is gauge invariant. This is the basis of the CSDR
scheme \cite{Manton},\cite{Review},\cite{Kuby}, which assumes
that $B$ is a compact coset space, $S/R$.

In the CSDR scheme one starts with a Yang-Mills-Dirac Lagrangian,
with gauge group $G$, defined on a
 $D$-dimensional spacetime $M^{D}$, with metric $g^{MN}$, which is compactified to $ M^{4}
\times S/R$ with $S/R$ a coset space. The metric is assumed to
have the form
\begin{equation}
g^{MN}=
\left[\begin{array}{cc}\eta^{\mu\nu}&0\\0&-g^{ab}\end{array}
\right],
\end{equation}
where $\eta^{\mu\nu}= diag(1,-1,-1,-1)$ and $g^{ab}$ is the coset
space metric. The requirement that transformations of the fields
under the action of the symmetry group of $S/R$ are compensated by
gauge transformations lead to certain constraints on the fields.
The solution of these constraints provides us with the
four-dimensional unconstrained fields as well as with the gauge
invariance that remains in the theory after dimensional reduction.
Therefore a potential unification of all low energy interactions,
gauge, Yukawa and Higgs is achieved, which was the first
motivation of this framework.

It is interesting to note that the fields obtained using the CSDR
approach are the first terms in the expansion of the
$D$-dimensional fields in harmonics of the internal space $B$ and
are massless after the first stage of the symmetry breaking which
is geometrical. The effective field theories resulting from
compactification of higher dimensional theories contain also
towers of massive higher harmonics (Kaluza-Klein) excitations,
whose contributions at the quantum level alter the behaviour of
the running couplings from logarithmic to power \cite{Taylor}. As
a result the traditional picture of unification of couplings may
change drastically \cite{Dienes}. Higher dimensional theories have
also been studied at the quantum level using the continuous Wilson
renormalization group \cite{Kubo} which can be formulated in any
number of space-time dimensions with results in agreement with the
treatment involving massive Kaluza-Klein excitations.

The group $S$ acts as a symmetry group on the the extra
coordinates. The CSDR scheme demands that an $S$-transformation of
the extra $d$ coordinates is a gauge transformation of the fields
that are defined on $M^{4}\times S/R$,  thus a gauge invariant
Lagrangian written on this space is independent of the extra
coordinates.

To see some of the details of the CDRS let us consider a
$D$-dimensional Yang-Mills-Dirac theory with gauge group $G$
defined on a manifold $M^{D}$ which as stated will be compactified
to $M^{4}\times S/R$, $D=4+d$, $d=dimS-dimR$:
\begin{equation}
A=\int d^{4}xd^{d}y\sqrt{-g}\Bigl[-\frac{1}{4}
Tr\left(F_{MN}F_{K\Lambda}\right)g^{MK}g^{N\Lambda}
+\frac{i}{2}\overline{\psi}\Gamma^{M}D_{M}\psi\Bigr] ,
\end{equation}
where
\begin{equation}
D_{M}= \partial_{M}-\theta_{M}-A_{M},
\end{equation}
with
\begin{equation}
\theta_{M}=\frac{1}{2}\theta_{MN\Lambda}\Sigma^{N\Lambda}
\end{equation}
the spin connection of $M^{D}$, and
\begin{equation}
F_{MN}
=\partial_{M}A_{N}-\partial_{N}A_{M}-\left[A_{M},A_{N}\right] ,
\end{equation}
where $M$, $N$ run over the $D$-dimensional space. The fields
$A_{M}$ and $\psi$ are, as explained, symmetric in the sense that
any transformation under symmetries of $S/R$  is compensated by
gauge transformations. The fermion fields can be in any
representation $F$ of $G$ unless a further symmetry such as
supersymmetry is required. So let $\xi_{A}^{\alpha}$, $A
=1,\ldots,dimS$, be the Killing vectors which generate the
symmetries of $S/R$ and $W_{A}$ the compensating gauge
transformation associated with $\xi_{A}$. Defining next the
infinitesimal coordinate transformation as $\delta_{A} \equiv
L_{\xi_{A}}$, i.e. the Lie derivative with respect to $\xi$, we
obtain the following consraints for the scalar,vector and spinor
fields,
\begin{eqnarray}
\delta_{A}\phi&=&\xi_{A}^{\alpha}\partial_{\alpha}\phi=D(W_{A})\phi,
\nonumber \\
\delta_{A}A_{\alpha}&=&\xi_{A}^{\beta}\partial_{\beta}A_{\alpha}+\partial_{\alpha}
\xi_{A}^{\beta}A_{\beta}=\partial_{\alpha}W_{A}-[W_{A},A_{\alpha}],\nonumber \\
\delta_{A}\psi&=&\xi_{A}^{\alpha}\psi-\frac{1}{2}G_{Abc}\Sigma^{bc}\psi=
D(W_{A})\psi.
\end{eqnarray}
$W_{A}$ depend only on internal coordinates $y$ and $D(W_{A})$
represents a gauge transformation in the appropriate
representation of the fields. $G_{Abc}$ represents a tangent space
rotation of the spinor fields. The variations $\delta_{A}$
satisfy, $[\delta_{A},\delta_{B}]=f_{AB}^{\\C}\delta_{C}$ and
lead to the following consistency relation for $W_{A}$'s,
\begin{equation}
\xi_{A}^{\alpha}\partial_{\alpha}W_{B}-\xi_{B}^{\alpha}\partial_{\alpha}
W_{A}-\left[W_{A},W_{B}\right]=f_{AB}^{\ \ C}W_{C}.
\end{equation}
 Furthermore the W's themselves transform under a gauge
transformation \cite{Review} as,
\begin{equation}
\widetilde{W_{A}}=gW_{A}g^{-1}+(\delta_{A}g)g^{-1}.
\end{equation}
Using eq.(8) and the fact that the Lagrangian is independent of
$y$ we can do all calculations at $y=0$ and choose a gauge where
$W_{a}=0$.

The detailed analysis of the constraints (6) given in
refs.\cite{Manton},\cite{Review} provides us with the
four-dimensional unconstrained fields as well as with the gauge
invariance that remains in the theory after dimensional reduction.
Here we present only the results. The components $A_{\mu}(x,y)$ of
the initial gauge field $A_{M}(x,y)$ become, after dimensional
reduction, the four-dimensional gauge fields and furthermore they
are independent of $y$. In addition one can find that they have to
commute with the elements of the $R_{G}$ subgroup of $G$. Thus the
four-dimensional gauge group $H$ is the centralizer of $R$ in $G$,
$H=C_{G}(R_{G})$. Similarly, the $A_{\alpha}(x,y)$ components of
$A_{M}(x,y)$ denoted by $\phi_{\alpha}(x,y)$ from now on, become
scalars at four dimensions. These fields transform under $R$ as a
vector $v$, i.e.
\begin{eqnarray}
S &\supset& R \nonumber \\
adjS &=& adjR+v.
\end{eqnarray}
Moreover $\phi_{\alpha}(x,y)$ act as an intertwining operator
connecting induced representations of $R$ acting on $G$ and $S/R$.
This implies, exploiting Schur's lemma, that the transformation
properties of the fields $\phi_{\alpha}(x,y)$ under $H$ can be
found if we express the adjoint representation of $G$ in terms of
$R_{G} \times H$ :
\begin{eqnarray}
G &\supset& R_{G} \times H \nonumber \\
 adjG &=&(adjR,1)+(1,adjH)+\sum(r_{i},h_{i}).
\end{eqnarray}
Then if $v=\sum s_{i}$, where each $s_{i}$ is an irreducible
representation of $R$, there survives an $h_{i}$ multiplet for
every pair $(r_{i},s_{i})$, where $r_{i}$ and $s_{i}$ are
identical irreducible representations of $R$.

Turning next to the fermion fields
\cite{Manton},\cite{Review},\cite{Chapline},\cite{Slansky},\cite{Palla},
similarly to scalars, they act as intertwining operators between
induced representations acting on $G$ and the tangent space of
$S/R$, $SO(d)$. Proceeding along similar lines as in the case of
scalars to obtain the representation of $H$ under which the four
dimensional fermions transform, we have to decompose the
representation $F$ of the initial gauge group in which the
fermions are assigned under $R_{G} \times H$, i.e.
\begin{equation}
F= \sum (t_{i},h_{i}),
\end{equation}
and the spinor of $SO(d)$ under $R$
\begin{equation}
\sigma_{d} = \sum \sigma_{j}.
\end{equation}
Then for each pair $t_{i}$ and $\sigma_{i}$, where $t_{i}$ and
$\sigma_{i}$ are identical irreducible representations there is an
$h_{i}$ multiplet of spinor fields in the four dimensional theory.
In order however  to obtain chiral fermions in the effective
theory we have to impose further requirements. We first impose the
Weyl condition in $D$ dimensions. In $D = 4n+2$ dimensions which
is the case at hand, the decomposition of the left handed, say
spinor under $SU(2) \times SU(2) \times SO(d)$ is
\begin{equation}
\sigma _{D} = (2,1,\sigma_{d}) + (1,2,\overline{\sigma}_{d}).
\end{equation}
So we have in this case the decompositions
\begin{equation}
\sigma_{d} = \sum \sigma_{k},~\overline{\sigma}_{d}= \sum
\overline{\sigma}_{k}.
\end{equation}
Let us start from a vector-like representation $F$ for the
fermions. In this case each term $(t_{i},h_{i})$ in eq.(11) will
be either self-conjugate or it will have a partner $(
\overline{t}_{i},\overline{h}_{i} )$. According to the rule
described in eqs.(11),(12) and considering $\sigma_{d}$ we will
have in four dimensions left-handed fermions transforming as $
f_{L} = \sum h^{L}_{k}$. It is important to notice that since
$\sigma_{d}$ is non self-conjugate, $f_{L}$ is non self-conjugate
too. Similarly from $\overline{\sigma}_{d}$ we will obtain the
right handed representation $ f_{R}= \sum \overline{h}^{R}_{k}$
but as we have assumed that $F$ is vector-like,
$\overline{h}^{R}_{k}\sim h^{L}_{k}$. Therefore there will appear
two sets of Weyl fermions with the same quantum numbers under $H$.
This is already a chiral theory, but still one can go further and
try to impose the Majorana condition in order to eliminate the
doubling of the fermionic spectrum. Clearly if we had started with
$F$ complex, we should have again a chiral theory since in this
case $\overline{h}^{R}_{k}$ is different from $h^{L}_{k}$
$(\sigma_{d}$ non self-conjugate). Nevertheless starting with $F$
vector-like is much more appealing and will be used in the
following along with the Majorana condition. The Majorana
condition can be imposed in $D = 2,3,4+8n$ dimensions and is given
by $\psi = C(\overline\psi)^{T}$, where $C$ is the $D$-dimensional
charge conjugation matrix. Majorana and Weyl conditions are
compatible in $D=4n+2$ dimensions. Then in our case if we start
with Weyl-Majorana spinors in $D=4n+2$ dimensions we force $f_{R}$
to be the charge conjugate to $f_{L}$, thus arriving in a theory
with fermions only in $f_{L}$. Furthermore if $F$ is to be real,
then we have to have $D=2+8n$, while for $F$ pseudoreal $D=6+8n$.

Starting with an anomaly free theory in higher dimensions, in
ref.\cite{Witten} was given the condition that has to be
fulfilled in order to obtain anomaly free theories in four
dimensions after dimensional reduction. The condition restricts
the allowed embeddings of $R$ into $G$ \cite{Pilch},\cite{Review}.
For $G=E_{8}$ in ten dimensions the condition takes the form
\begin{equation}
l(G) = 60,
\end{equation}
where $l(G)$ is the sum over all indices of the $R$
representations appearing in the decomposition of the $248$
representation of $E_{8}$ under $ E_{8} \supset R \times H$. The
normalization is such that the vector representation in eq.(9)
which defines the embedding of $R$ into $SO(6)$, has index two.

Next let us obtain the four-dimensional effective action. Assuming
that the metric is block diagonal, taking into account all the
constraints and integrating out the extra coordinates we obtain in
four dimensions the following Lagrangian :
\begin{equation}
A=C \int d^{4}x \biggl( -\frac{1}{4} F^{t}_{\mu
\nu}{F^{t}}^{\mu\nu}+\frac{1}{2}(D_{\mu}\phi_{\alpha})^{t}
(D^{\mu}\phi^{\alpha})^{t}
+V(\phi)+\frac{i}{2}\overline{\psi}\Gamma^{\mu}D_{\mu}\psi-\frac{i}{2}
\overline{\psi}\Gamma^{a}D_{a}\psi\biggr),
\end{equation}
where $D_{\mu} = \partial_{\mu} - A_{\mu}$ and $D_{a}=
\partial_{a}- \theta_{a}-\phi_{a}$ with  $\theta_{a}=
\frac{1}{2}\theta_{abc}\Sigma^{bc}$ the connection of the coset
space, while $C$ is the volume of the coset space. The potential
$V(\phi)$ is given by
\begin{equation}
V(\phi) = - \frac{1}{4} g^{ac}g^{bd}Tr( f _{ab}^{C}\phi_{C} -
[\phi_{a},\phi_{b}] ) (f_{cd}^{D}\phi_{D} - [\phi_{c},\phi_{d}] )
,
\end{equation}
where, $A=1,\ldots,dimS$ and $f$ ' s are the structure constants
appearing in the commutators of the generators of the Lie algebra
of S. The expression (17) for $V(\phi)$ is only formal because
$\phi_{a}$ must satisfy the constraints coming from eq.(6),
\begin{equation}
f_{ai}^{D}\phi_{D} - [\phi_{a},\phi_{i}] = 0,
\end{equation}
where the $\phi_{i}$ generate $R_{G}$. These constraints imply
that some components $\phi_{a}$'s are zero, some are constants and
the rest can be identified with the genuine Higgs fields. When
$V(\phi)$ is expressed in terms of the unconstrained independent
Higgs fields, it remains a quartic polynomial which is invariant
under gauge transformations of the final gauge group $H$, and its
minimum determines the vacuum expectation values of the Higgs
fields \cite{Harnad},\cite{Farakos}.

In the fermion part of the Lagrangian the first term is just the
kinetic term of fermions, while the second is the Yukawa term
\cite{Kapetanakis}. Note that since $\psi$ is a Majorana-Weyl
spinor in ten dimensions the representation in which the fermions
are assigned under the gauge group must be real. The last term in
eq.(16) can be written as
\begin{equation}
L_{Y}= -\frac{i}{2}\overline{\psi}\Gamma^{a}(\partial_{a}-
\frac{1}{2}f_{ibc}e^{i}_{\gamma}e^{\gamma}_{a}\Sigma^{bc}-
\frac{1}{2}G_{abc}\Sigma^{bc}- \phi_{a}) \psi \nonumber \\
=\frac{i}{2}\overline{\psi}\Gamma^{a}\nabla_{a}\psi+
\overline{\psi}V\psi ,
\end{equation}
where
\begin{eqnarray}
\nabla_{a}& =& - \partial_{a} +
\frac{1}{2}f_{ibc}e^{i}_{\gamma}e^{\gamma}_{a}\Sigma^{bc} + \phi_{a},\\
 V&=&\frac{i}{4}\Gamma^{a}G_{abc}\Sigma^{bc},
\end{eqnarray}
and we have used the full connection with torsion \cite{Review}
given by
\begin{equation}
\theta_{\ \ c b}^{a} = - f_{\ \
ib}^{a}e^{i}_{\alpha}e^{\alpha}_{c}-(D_{\ \ cb}^{a} +
\frac{1}{2}\Sigma_{\ \ cb}^{a}) = - f_{\ \
ib}^{a}e^{i}_{\alpha}e^{\alpha}_{c}- G_{\ \ cb}^{a}
\end{equation}
with
\begin{equation}
D_{\ \ cb}^{a} = g^{ad}\frac{1}{2}[f_{db}^{\ \ e}g_{ec} + f_{
cb}^{\ \ e}g_{de} - f_{cd}^{\ \ e}g_{be}]
\end{equation}
and
\begin{equation}
\Sigma_{abc}= 2\tau(D_{abc} +D_{bca} - D_{cba}).
\end{equation}
 We have already noticed that the CSDR constraints tell us that
$\partial_{a}\psi= 0$. Furthermore we can consider the Lagrangian
at the point $y=0$, due to its invariance under
$S$-transformations, and as we mentioned $e^{i}_{\gamma}=0$ at
that point. Therefore eq.(20) becomes just $\nabla_{a}= \phi_{a}$
and the term $\frac{i}{2}\overline{\psi}\Gamma^{a}\nabla_{a}\psi $
in eq.(19) is exactly the Yukawa term.

Let us examine now the last term appearing in eq.(19). One can
show easily that the operator $V$ anticommutes with the
six-dimensional helicity operator \cite{Review}. Furthermore one
can show that $V$ commutes with the $T_{i}=
-\frac{1}{2}f_{ibc}\Sigma^{bc}$ ($T_{i}$ close the $R$-subalgebra
of $SO(6)$). In turn we can draw the conclusion, exploiting
Schur's lemma, that the non-vanishing elements of $V$ are only
those which appear in the decomposition of both $SO(6)$ irreps
$4$ and $\overline{4}$, e.g. the singlets. Since this term is of
pure geometric nature, we reach the conclusion that the singlets
in $4$ and $\overline{4}$ will acquire large geometrical masses,
a fact that has serious phenomenological implications. In
supersymmetric theories defined in higher dimensions, it means
that the gauginos obtained in four dimensions after dimensional
reduction receive masses comparable to the compactification
scale. However as we shall see in the next sections this result
changes in presence of torsion. We note that for symmetric coset
spaces the $V$ operator is absent since in that case $f_{ab}^{c}$
vanish by definition.
\section{Soft Supersymmetry Breaking by Dimensional Reduction over Non-Symmetric
 Coset Spaces.}
 Recently a lot of interest has been triggered by the
possibility that superstrings can be defined at the TeV scale
\cite{Anton}. The string tension became an arbitrary parameter and
can be anywhere below the Planck scale and as low as TeV. The main
advantage of having the string tension at the TeV scale, besides
the obvious experimental interest, is that it offers an automatic
protection to the gauge hierarchy \cite{Anton}, alternative to low
energy supersymmetry \cite{Dim}, or dynamical electroweak symmetry
breaking \cite{Fahri},\cite{Marciano},\cite{Trianta}. However the
only vacua of string theory free of all pathologies are
supersymmetric. Then the original supersymmetry of the theory, not
being necessary in four dimensions, could be broken by the
dimensional reduction procedure.

The weakly coupled ten-dimensional $E_{8} \times E_{8}$
supersymmetric gauge theory is one of the few to posses  the
advantage of anomaly freedom \cite{Green} and has been extensively
used in efforts to describe quantum gravity along with the
observed low energy interactions in the heterotic string framework
\cite{Theisen}. In addition its strong coupling limit  provides an
interesting example of the realization of the brane picture, i.e.
$E_{8}$ gauge fields and matter live on the two 10-dimensional
boundaries, while gravitons propagate in the eleven-dimensional
bulk \cite{Horava}.

In the following sections we shall be reducing a supersymmetric
ten-dimensional gauge theory based on $E_{8}$ over the
six-dimensional coset spaces $Sp(4)/(SU(2) \times
U(1))_{non-max.}$ and $SU(3)/U(1) \times U(1)$ and examine the
consequences of the resulting four-dimensional theory mostly as
far as supersymmetry breaking is concerned.
\subsection{{\it Supersymmerty
breaking by dimensional reduction over \\ $Sp(4)/(SU(2) \times
U(1))_{non-max.}$}} In the present study we start with a
ten-dimensional supersymmetric gauge theory based on the group
$E_{8}$ and reduce it over the non-symmetric coset $Sp(4)/(SU(2)
\times U(1))_{non-max.}$. Therefore in the terminology of section
2 we have chosen $G=E_{8}$, $B=Sp(4)/(SU(2) \times
U(1))_{non-max.}$, $D=10$ and Weyl-Majorana fermions belonging in
the adjoint of $G$. We start by giving the decompositions to be
used,
$$E_{8} \supset SU(3) \supset SU(2) \times U(1) \times E_{6}.$$
The decomposition of $248$ of $E_{8}$ under $SU(3) \times E_{6}$
is given by $$ 248 = (8,1) + (1,78) + (3,27) +
(\overline{3},\overline{27}),$$ while under $ (SU(2) \times U(1))
\times E_{6}$ is the following:
\begin{eqnarray}
248 = (3_{0},1)+(1_{0},1)+(1_{0},78)+(2_{3},1)+(2_{-3},1)\nonumber\\
+(2_{1},27)+(2_{-1},\overline{27})+(1_{-2},27)
+(1_{2},\overline{27}).
\end{eqnarray}
In the present case $R$ is chosen to be identified with the $SU(2)
\times U(1)$ of the latter of the above decompositions. Therefore
the resulting four-dimensional gauge theory is based on the group
$$H=C_{E_{8}}(SU(2) \times U(1))= E_{6} \times U(1),$$
where the $U(1)$ appears since the $U(1)$ in $R$ centralizes
itself. The $R=SU(2) \times U(1)$ content of $Sp(4)/(SU(2) \times
U(1))_{non-max.}$ vector and spinor are
$2_{1}+2_{-1}+1_{2}+1_{-2}$ and $2_{1}+1_{0}+1_{-2}$
respectively. Thus applying the CSDR rules (9),(10) and (11),(12)
we find that the surviving fields in four dimensions can be
organized in a ${\cal N}=1$ vector supermultiplet $V^{\alpha}$
which transforms as $78$ of $E_{6}$, a ${\cal N}=1$ $U(1)$ vector
supermultiplet $V$ and chiral supermultiplets ($B^{i}$, $C^{i}$),
transforming as $(27,1)$, and $(27,-2)$ under the gauge group
$E_{6} \times U(1)$.

We find that the potential of the four-dimensional theory, in
terms of the physical scalar fields $\beta^{i}$, and $\gamma^{i}$
is given by
\begin{eqnarray}
V(\beta^{i}, \gamma^{j})= const
-\frac{6}{R_{1}^{2}}\beta^{i}\beta_{i}
-\frac{4}{R_{2}^{2}}\gamma^{i}\gamma_{i} \nonumber\\
+\bigg[4\sqrt{\frac{10}{7}}R_{2}
\bigl(\frac{1}{R_{2}^{2}}+\frac{1}{2R_{1}^{2}}\bigr)
d_{ijk}\beta^{i}\beta^{j}\gamma^{k} + h.c \biggr] \nonumber \\
+6\biggl(\beta^{i}(G^{\alpha})_{i}^{j}\beta_{j}
+\gamma^{i}(G^{\alpha})_{i}^{j}\gamma_{j}\biggr)^{2}\nonumber\\
+\frac{1}{3}\biggl(\beta^{i}(1\delta_{i}^{j})\beta_{j}+
\gamma^{i}(-2\delta_{i}^{j})\gamma_{j}\biggr)^{2}\nonumber\\
+\frac{5}{7}\beta^{i}\beta^{j}d_{ijk}d^{klm}\beta_{l}\beta_{m}
+4\frac{5}{7}\beta^{i}\gamma^{j}d_{ijk}d^{klm}\beta_{l}\gamma_{m}.
\end{eqnarray}
From the potential (26) we can determine the $F$-, $D$- and  the
scalar soft terms which break softly the supersymmetric theory
obtained by CSDR over $Sp(4)/(SU(2) \times U(1))_{non-max.}$.
Specifically we find that the $F$-term contributions to the
potential (26) come from the superpotential
\begin{equation}
{\cal W}(B^{i},C^{j})= \sqrt{\frac{5}{7}}d_{ijk}B^{i}B^{j}C^{k}.
\end{equation}
Similarly the $D$-term contributions to the potential (26) are
given by the sum
\begin{equation}
\frac{1}{2}D^{\alpha}D^{\alpha}+\frac{1}{2}DD,
\end{equation}
where
$$D^{\alpha}=\sqrt{12}\bigl(\beta^{i}(G^{\alpha})_{i}^{j}\beta_{j}
+\gamma^{i}(G^{\alpha})_{i}^{j}\gamma_{j}\bigr)$$ and $$
D=\sqrt{\frac{2}{3}}\bigl(\beta^{i}(1\delta_{i}^{j})\beta_{j}+
\gamma^{i}(-2\delta_{i}^{j})\gamma_{j}\bigr)$$ corresponding to
the vector supermultiplets of $E_{6} \times U(1)$. The remaining
terms in the potential (26) are the soft breaking mass and
trilinear terms and they form the scalar SSB part of the
Lagrangian,
\begin{equation}
{ \cal L}_{scalarSSB}= -\frac{6}{R_{1}^{2}}\beta^{i}\beta_{i}
-\frac{4}{R_{2}^{2}}\gamma^{i}\gamma_{i} +
\bigg[4\sqrt{\frac{10}{7}}R_{2}
\bigl(\frac{1}{R_{2}^{2}}+\frac{1}{2R_{1}^{2}}\bigr)
d_{ijk}\beta^{i}\beta^{j}\gamma^{k} + h.c \biggr].
\end{equation}
 The gaugino mass has been calculated in ref.\cite{Kapetanakis} to be
\begin{equation}
M=(1+3\tau)\frac{R_{2}^{2}+2R_{1}^{2}}{8R_{1}^{2}R_{2}}.
\end{equation}
We note that the chosen embedding of $R=SU(2) \times U(1)$ in
$E_{8}$ satisfies the condition (15) which guarantees the
renormalizability of the four dimensional theory, while the
absence of any other term that does not belong to the
supersymmetric $E_{6} \times U(1)$ theory or to its SSB sector
guarantees the improved ultraviolet behaviour of the theory.
Finally note the contribution of the torsion in the gaugino mass
(30).
\subsection{{\it Supersymmetry breaking by reduction over $SU(3)/(U(1)
\times U(1))$.}} In this model the only difference as compared to
the previous one is that the chosen coset space to reduce the
same theory is the non-symmetric $B=SU(3)/U(1) \times U(1)$. The
decompositions to be used are $$ E_{8} \supset SU(2) \times U(1)
\times E_{6} \supset U(1)_{1} \times U(1)_{2} \times E_{6}. $$ The
$248$ of $E_{8}$ is decomposed under $SU(2) \times U(1)$
according to (25) whereas the decomposition under $U(1)_{1} \times
U(1)_{2}$ is the following:
\begin{eqnarray}
 248 = (0,0;1)+(0,0;1)+(3,\frac{1}{2};1)+(-3,\frac{1}{2};1)+\nonumber\\
(0,-1;1)+(0,1;1)+(-3,-\frac{1}{2};1)+(-3,-\frac{1}{2};1)+\nonumber\\
(0,0;78)+(3,\frac{1}{2};27)+(-3,\frac{1}{2};27)+(0,-1;27)+\nonumber\\
(-3,-\frac{1}{2};\overline{27})+(3,-\frac{1}{2};\overline{27})
+(0,1;\overline{27}).
\end{eqnarray}
In the present case $R$ is chosen to be identified with the
$U(1)_{1} \times U(1)_{2}$ of the latter decomposition. Therefore
the resulting four-dimensional gauge group is $$
H=C_{E_{8}}(U(1)_{1} \times U(1)_{2}) = U(1)_{1} \times U(1)_{2}
\times E_{6}. $$ Again the two $U(1)$'s appear because $R\
(=U(1)_{1} \times U(1)_{2})$ centralizes itself. The $R=U(1)
\times U(1)$ content of $SU(3)/U(1) \times U(1)$ vector and spinor
are $(3,\frac{1}{2})+(-3,\frac{1}{2})
+(0,-1)+(-3,-\frac{1}{2})+(3,-\frac{1}{2})+(0,1)$ and
$(0,0)+(3,\frac{1}{2})+(-3,\frac{1}{2}) +(0,-1)$ respectively.
Thus applying the CSDR rules (9) -- (12) we find that the
surviving fields in four dimensions are three ${\cal N}=1$ vector
multiplets $V^{\alpha},V_{(1)},V_{(2)}$, (where $\alpha$ is an
$E_{6}$, $78$ index and the other two refer to the two $U(1)'s$)
containing the gauge fields of $U(1)_{1} \times U(1)_{2} \times
E_{6}$. The matter content consists of three ${\cal N}=1$ chiral
multiplets ($A^{i}$, $B^{i}$, $C^{i}$) with $i$ an $E_{6}$, $27$
index and three ${\cal N}=1$ chiral multiplets ($A$, $B$, $C$)
which are $E_{6}$ singlets and carry $U(1)_{1} \times U(1)_{2}$
charges.

We find that the unconstrained  scalar fields transform under
$U(1)_{1} \times U(1)_{2} \times E_{6}$ as
\begin{equation}
\alpha_{i} \sim (3,\frac{1}{2};27),\alpha \sim (3,\frac{1}{2};1),
\beta_{i} \sim (-3,\frac{1}{2};27),\beta \sim (-3,\frac{1}{2};1),
\gamma_{i} \sim (0,-1;27),\gamma \sim (0,-1;1).
\end{equation}
The potential of the four dimensional theory in terms of the
unconstrained fields given in (32) is the following
\begin{eqnarray}
V(\alpha^{i},\beta^{j},\gamma^{k},\alpha,\beta,\gamma)= const. +
\biggl( \frac{4R_{1}^{2}}{R_{2}^{2}R_{3}^{2}}-\frac{8}{R_{1}^{2}}
\biggr)\alpha^{i}\alpha_{i} +\biggl(
\frac{4R_{1}^{2}}{R_{2}^{2}R_{3}^{2}}-\frac{8}{R_{1}^{2}}
\biggr)\overline{\alpha}\alpha \nonumber \\
+\biggl(\frac{4R_{2}^{2}}{R_{1}^{2}R_{3}^{2}}-\frac{8}{R_{2}^{2}}\biggr)
\beta^{i}\beta_{i}
+\biggl(\frac{4R_{2}^{2}}{R_{1}^{2}R_{3}^{2}}-\frac{8}{R_{2}^{2}}\biggr)
\overline{\beta}\beta \nonumber \\
+\biggl(\frac{4R_{3}^{2}}{R_{1}^{2}R_{2}^{2}}
-\frac{8}{R_{3}^{2}}\biggr)\gamma^{i}\gamma_{i}
+\biggl(\frac{4R_{3}^{2}}{R_{1}^{2}R_{2}^{2}}
-\frac{8}{R_{3}^{2}}\biggr)\overline{\gamma}\gamma \nonumber\\
+\biggl[\sqrt{2}80\biggl(\frac{R_{1}}{R_{2}R_{3}}+\frac{R_{2}}{R_{1}
R_{3}}+\frac{R_{3}}{R_{2}R_{1}}\biggr)d_{ijk}\alpha^{i}\beta^{j}\gamma^{k}\nonumber\\
+\sqrt{2}80\biggl(\frac{R_{1}}{R_{2}R_{3}}+\frac{R_{2}}{R_{1}
R_{3}}+\frac{R_{3}}{R_{2}R_{1}}\biggr)\alpha\beta\gamma+
h.c\biggr]\nonumber\\
+\frac{1}{6}\biggl(\alpha^{i}(G^{\alpha})_{i}^{j}\alpha_{j}
+\beta^{i}(G^{\alpha})_{i}^{j}\beta_{j}
+\gamma^{i}(G^{\alpha})_{i}^{j}\gamma_{j}\biggr)^{2}\nonumber\\
+\frac{10}{6}\biggl(\alpha^{i}(3\delta_{i}^{j})\alpha_{j} +
\overline{\alpha}(3)\alpha + \beta^{i}(-3\delta_{i}^{j})\beta_{j}
+ \overline{\beta}(-3)\beta \biggr)^{2}\nonumber \\
+\frac{40}{6}\biggl(\alpha^{i}(\frac{1}{2}\delta_{i}^{j})\alpha_{j}
+ \overline{\alpha}(\frac{1}{2})\alpha +
\beta^{i}(\frac{1}{2}\delta^{j}_{i})\beta_{j} +
\overline{\beta}(\frac{1}{2})\beta +
\gamma^{i}(-1\delta_{i}^{j})\gamma_{j} +
\overline{\gamma}(-1)\gamma \biggr)^{2}\nonumber \\
+40\alpha^{i}\beta^{j}d_{ijk}d^{klm}\alpha_{l}\beta_{m}
+40\beta^{i}\gamma^{j}d_{ijk}d^{klm}\beta_{l}\gamma_{m}
+40\alpha^{i}\gamma^{j}d_{ijk}d^{klm}\alpha_{l}\gamma_{m}\nonumber\\
+40(\overline{\alpha}\overline{\beta})(\alpha\beta) +
40(\overline{\beta}\overline{\gamma})(\beta\gamma) +
40(\overline{\gamma}\overline{\alpha})(\gamma\alpha).
\end{eqnarray}
From the potential (33) we read the $F$-, $D$- and scalar soft
terms. The $F$-terms are obtained from the superpotential
\begin{equation}
{ \cal W}(A^{i},B^{j},C^{k},A,B,C)
=\sqrt{40}d_{ijk}A^{i}B^{j}C^{k} + \sqrt{40}ABC.
\end{equation}
The $D$-terms have the structure
\begin{equation}
\frac{1}{2}D^{\alpha}D^{\alpha}+\frac{1}{2}D_{1}D_{1}+\frac{1}{2}D_{2}D_{2},
\end{equation}
where $$D^{\alpha}= \frac{1}{\sqrt{3}}
\biggl(\alpha^{i}(G^{\alpha})_{i}^{j}\alpha_{j}
+\beta^{i}(G^{\alpha})_{i}^{j}\beta_{j}
+\gamma^{i}(G^{\alpha})_{i}^{j}\gamma_{j}\biggr) ,$$ $$D_{1}=
\sqrt{ \frac{10}{3} }\biggl(\alpha^{i}(3\delta_{i}^{j})\alpha_{j}
+ \overline{\alpha}(3)\alpha +
\beta^{i}(-3\delta_{i}^{j})\beta_{j} + \overline{\beta}(-3)\beta
\biggr)$$ and $$D_{2} = \sqrt{ \frac{40}{3}
}\biggl(\alpha^{i}(\frac{1}{2}\delta_{i}^{j})\alpha_{j} +
\overline{\alpha}(\frac{1}{2})\alpha +
\beta^{i}(\frac{1}{2}\delta^{j}_{i})\beta_{j} +
\overline{\beta}(\frac{1}{2})\beta +
\gamma^{i}(-1\delta_{i}^{j})\gamma_{j} +
\overline{\gamma}(-1)\gamma \biggr) ,$$ which correspond to the
$U(1)_{1} \times U(1)_{2} \times E_{6}$ vector supermultiplet
content of the four-dimensional theory. The rest terms are the
trilinear and mass terms which break supersymmetry softly and
they form the scalar SSB part of the Lagrangian,
\begin{eqnarray}
\lefteqn{{\cal L}_{scalarSSB}=  \biggl(
\frac{4R_{1}^{2}}{R_{2}^{2}R_{3}^{2}}-\frac{8}{R_{1}^{2}}
\biggr)\alpha^{i}\alpha_{i} +\biggl(
\frac{4R_{1}^{2}}{R_{2}^{2}R_{3}^{2}}-\frac{8}{R_{1}^{2}}
\biggr)\overline{\alpha}\alpha} \nonumber\\
& &
+\biggl(\frac{4R_{2}^{2}}{R_{1}^{2}R_{3}^{2}}-\frac{8}{R_{2}^{2}}\biggr)
\beta^{i}\beta_{i}
+\biggl(\frac{4R_{2}^{2}}{R_{1}^{2}R_{3}^{2}}-\frac{8}{R_{2}^{2}}\biggr)
\overline{\beta}\beta
+\biggl(\frac{4R_{3}^{2}}{R_{1}^{2}R_{2}^{2}}
-\frac{8}{R_{3}^{2}}\biggr)\gamma^{i}\gamma_{i}
+\biggl(\frac{4R_{3}^{2}}{R_{1}^{2}R_{2}^{2}}
-\frac{8}{R_{3}^{2}}\biggr)\overline{\gamma}\gamma \nonumber\\
& &
+\biggl[\sqrt{2}80\biggl(\frac{R_{1}}{R_{2}R_{3}}+\frac{R_{2}}{R_{1}
R_{3}}+\frac{R_{3}}{R_{2}R_{1}}\biggr)d_{ijk}\alpha^{i}\beta^{j}\gamma^{k}
\nonumber \\
& &+\sqrt{2}80\biggl(\frac{R_{1}}{R_{2}R_{3}}+\frac{R_{2}}{R_{1}
R_{3}}+\frac{R_{3}}{R_{2}R_{1}}\biggr)\alpha\beta\gamma+
h.c\biggr].
\end{eqnarray}

Finally in order to determine the gaugino mass we calculate the V
operator appearing in eq.(19). We find that the gauginos acquire a
geometrical mass
\begin{equation}
M=(1+3\tau)\frac{(R_{1}^{2}+R_{2}^{2}+R_{3}^{2})}{8\sqrt{R_{1}^{2}R_{2}^{2}R_{3}^{2}}}.
\end{equation}
We note again that the chosen embedding satisfies the condition
(15) and the absence in the four-dimensional theory of any other
term that does not belong to the supersymmetric $U(1)_{1} \times
U(1)_{2} \times E_{6}$ gauge theory or to its SSB sector. The
gaugino mass (37) has a contribution from the torsion of the coset
space similarly to the reduction over the other non-symmetric
spaces. Contrary to the gaugino mass term the soft scalar terms of
the SSB do not receive contributions from the torsion in all
models. This is due to the fact that gauge fields, contrary to
fermions, do not couple to torsion.
\section{Conclusions}
 The CSDR was originally introduced as a scheme which, making use
 of higher dimensions, incorporates in a
unified manner the gauge and the {\it ad hoc} Higgs sector of the
spontaneously broken gauge theories in four dimensions
\cite{Manton}. Next fermions were introduced in the scheme and the
{\it ad hoc} Yukawa interactions have also been included in the
unified description \cite{Slansky}.

Considerable progress has also been made in attempts to describe
the observed low-energy world within the CSDR framework. Among the
new possibilities emerged from the subsequent studies of the CSDR
scheme are the following: (a) The possibility to obtain chiral
fermions in four dimensions resulting from vector-like
representations of the higher dimensional gauge theory
\cite{Chapline},\cite{Review}. This possibility can be realized
due the presence of non-trivial background gauge configurations
which are introduced by the CSDR constructions \cite{Salam}, (b)
The possibility to deform the metric of certain non-symmetric
coset spaces and thereby obtain more than one scales
\cite{Farakos},\cite{Review},\cite{Hanlon}, (c) The possibility to
use coset spaces, which are multiply connected. This can be
achieved by exploiting the discrete symmetries of the $S/R$
\cite{Kozimirov},\cite{Review}. Then one might introduce
topologically non-trivial gauge field \cite{Zoupanos}
configurations with vanishing field strength and induce additional
breaking of the gauge symmetry. It is the Hosotani mechanism
\cite{Hosotani} applied in the CSDR.

In the above list recently has been added the interesting
possibility that the popular softly broken supersymmetric four
dimensional chiral gauge theories might have their origin in a
higher dimensional supersymmetric theory with only vector
supermultiplet \cite{Pman}, which is dimensionally reduced over
non-symmetric coset spaces. In the present work we have extended
the previous observations
\cite{Review},\cite{Chapline},\cite{Kapetanakis} and the concrete
proposal of ref.\cite{Pman} in the remaining six-dimensional
non-symmetric coset spaces, demonstrating in this way that the
claim of ref.\cite{Pman} holds more generally and it is not just a
peculiarity of the coset space that was used.

Given the recent interest on the Scherk-Schwarz mechanism
\cite{Kawamura}, it is worth adding few comments concerning the
relation among the Scherk-Schwarz and our mechanism. Without
making any attempt to cover the many aspects of the subject
discussed over years it seems that Scherk and Schwarz
\cite{Scherk-Schwarz}, were influenced by the work of Forgacs and
Manton \cite{Manton}, which was done few months earlier and used
the generalized reduction on which the CSDR is based on, i.e. they
also allowed dependence of various fields on the compact space
coordinates corresponding to a gauge transformation. Moreover in
ref.\cite{Scherk-Schwarz}, among others, they have examined the
reduction of supersymmetric Yang-Mills theories in the above sense
as we do. The real difference is that they did the reduction on a
group manifold instead of coset space, which is a limiting case of
coset space with $R=I$ and has the obvious problem that the
resulting four-dimensional theory has no chiral fermions. They
claimed without going in the details that supersymmetry was
broken.

Schwarz more than twenty years later, in ref.\cite{Joel}, was
describing the basic idea of the Scherk-Schwarz mechanism as
follows: ``The idea is that in a theory with extra dimensions and
global symmetries that do not commute with supersymmetry ( $R$
symmetries and $(-1)^F$ are examples ), one could arrange for a
twisted compactification, and that this would break
supersymmetry." In case of ordinary reduction of a ten-dimensional
supersymmetric Yang-Mills  theory one obtains ${ \cal N}=4$
supersymmetric Yang-Mills theory in four dimensions. This has a
global SU(4) $R$ symmetry which is identified with the tangent
space $SO(6)$. In the CSDR in order to solve the constraints
imposed on the fermions one has to embed $R$ (of $S/R$) into
$SO(6)$. Moreover the four-dimensional Lagrangian resulting from
CSDR has an a global symmetry $R$ (of the $S/R$). Therefore the
CSDR satisfies automatically the criterion stated by Schwarz
above that could lead to supersymmetry breaking.
\section*{Acknowledgments}
We would like to thank L.~Alvarez-Gaume, C.~Bachas, G.~L.~Cardoso,
P.~Forgacs, A.~Kehagias, C.~Kounnas, G.~Koutsoumbas, D.~Luest and
D.~Suematsu for useful discussions.

\end{document}